\documentclass[twocolumn]{revtex4-1}
\usepackage{graphicx}
\usepackage{amsmath}
\usepackage{amsfonts}%
\usepackage{amssymb}
\usepackage{mathrsfs}
\usepackage{dcolumn}
\usepackage{bm}
\usepackage{float}
\usepackage{epsfig}
\usepackage{color}

\begin{document}
\title{Laser beam self-symmetrization in air in the multifilamentation regime}
\author{C. Mili\'{a}n}
\email[]{carles.milian@cpht.polytechnique.fr}
\affiliation{Centre de Physique Th\'{e}orique, CNRS, \'{E}cole Polytechnique, F-91128 Palaiseau, France}
\author{V. Jukna}
\affiliation{Centre de Physique Th\'{e}orique, CNRS, \'{E}cole Polytechnique, F-91128 Palaiseau, France}
\author{A. Couairon}
\affiliation{Centre de Physique Th\'{e}orique, CNRS, \'{E}cole Polytechnique, F-91128 Palaiseau, France}
\author{A. Houard}
\affiliation{Laboratoire d'Optique Appliqu\'{e}e, ENSTA ParisTech, \'{E}cole Polytechnique, CNRS, F-91761 Palaiseau, France}
\author{B. Forestier}
\affiliation{Laboratoire d'Optique Appliqu\'{e}e, ENSTA ParisTech, \'{E}cole Polytechnique, CNRS, F-91761 Palaiseau, France}
\author{J. Carbonnel}
\affiliation{Laboratoire d'Optique Appliqu\'{e}e, ENSTA ParisTech, \'{E}cole Polytechnique, CNRS, F-91761 Palaiseau, France}
\author{Y. Liu}
\affiliation{Laboratoire d'Optique Appliqu\'{e}e, ENSTA ParisTech, \'{E}cole Polytechnique, CNRS, F-91761 Palaiseau, France}
\author{B. Prade}
\affiliation{Laboratoire d'Optique Appliqu\'{e}e, ENSTA ParisTech, \'{E}cole Polytechnique, CNRS, F-91761 Palaiseau, France}
\author{A. Mysyrowicz}
\affiliation{Laboratoire d'Optique Appliqu\'{e}e, ENSTA ParisTech, \'{E}cole Polytechnique, CNRS, F-91761 
Palaiseau, France}

\begin{abstract}
We show experimental and numerical evidence of spontaneous self-symmetrization of focused laser beams experiencing multi-filamentation in air. The symmetrization effect is observed as the multiple filaments generated prior to focus approach the focal volume. This phenomenon is attributed to the nonlinear interactions amongst the different parts of the beam mediated by the optical Kerr effect, which leads to a symmetric redistribution of the wave vectors even when the beam consists of a bundle of many filaments.
\end{abstract}

\maketitle

\section{Introduction}
An intense laser pulse propagating in air undergoes various self induced transformations if its peak power exceeds a critical value \cite{MarburgerPQE75} $P_{cr}
\approx 4.7$ GW. 
The optical Kerr effect is responsible for an increase of the refractive index leading to beam self focusing, a cumulative process evolving towards a collapse.
An electron plasma is efficiently generated via optical field ionization when the beam intensity has increased above $\sim 10^{13}$ W/cm$^2$. Ionization is associated with nonlinear absorption of laser energy and with a decrease of the local refractive index. Both effects act against the growth of intensity by self-focusing and eventually arrest the collapse. The interplay between nonlinear effects leads to the formation of a narrow light filament leaving a plasma channel in its wake, surrounded by a laser energy reservoir. This reservoir maintains an energy flux toward the filament core that compensates for nonlinear absorption. This optical entity can live over several meters and be generated at long distances 
\cite{MechainOC05,CouaironPR07}. Ionization in the atmosphere by filamentation has been reported at a distance of $1$ km from the laser \cite{DurandOE13}. When the reservoir is exhausted and fails to feed the filament efficiently, a slowly diverging \textit{bright channel} is observed and can be detected at several kilometers away from the laser source. Several teams have reported an effect called beam self-cleaning in the low power regime $P \approx P_{cr}$ \cite{MollPRL03,PradeOL06,ChinAPB07,LiuPRA2007,heinsAPB13}. Moll and Gaeta showed that slightly elliptical beams with power close to $P_{cr}$ tend to increase their roundness and \textit{recover} a circularly symmetric profile when they undergo self-focusing, just before ionization. This effect has been interpreted as due to beam collapse and reshaping into a universal self-similar spatial profile, the Townes mode \cite{ChiaoPRL64,MollPRL03}, which is intrinsically spatially symmetric. Prade et al. \cite{PradeOL06} measured conical emission for ultraviolet pulses undergoing filamentation in air and reported self-improvement
of the beams's spatial mode quality, suggesting the generation of propagation invariant modes called nonlinear X-waves \cite{ContiPRL03,KolesikPRL04}. Liu and Chin have shown that the beam cleaning process only occurs for the frequency downshifted components of the supercontinuum generated during the filamentation process. They attributed this effect to the role of self-focusing as a spatial filter, filtering out the high-order mode components of the beam whereas the fundamental mode is focused, producing a high quality filament core. Naturally this process is promoted for redshifted frequencies which are generated in the leading edge of the pulse and undergo mainly self-focusing and diffraction \cite{LiuPRA2007}. From these examples, it is clear that even if beam self-cleaning received different interpretations, this effect occurs for beam powers moderately above the critical power.

For $P\gg P_{cr}$, however, a beam rises multiple filaments across the profile, as manifestations of the dynamics following local collapse. These filaments are born on beam intensity fluctuations \cite{MechainPRL04,MajusPRA09} 
and grow due to modulation instability \cite{ZakharovPD2009,BespalovJETPL66}  (MI). Several filament interactions manifested as cross phase modulation  (XPM), fusion, repulsion, spiraling, and fission  have been reported in this high power regime \cite{MlejnekPRL99,TzortzakisPRL01a,MechainPRL04,XiPRL06,KiranPRA10}. It has also been shown recently that focused multi-filamented beams maintain a substructure during the focal region and may produce thick filaments associated to unusually high intensity and plasma levels, coined as superfilaments \cite{PointPRL14}. Up to date, all studies done with high power beams focus on the filament features, rather than on the whole beam dynamics.

In this work we investigate an effect of the global structure of powerful ($P\gg P_{cr}$) beam profiles that remained unexplored. We demonstrate experimentally and by means of numerical simulations that under focusing conditions, powerful beams undergo self-induced symmetrization in the global scale above certain threshold power, $P_{th}\gg P_{cr}$, even if hundreds of filaments are born along propagation. As a consequence, the beam washes out all interactions and distortions it might have suffered prior to the focus. However the nature of this very robust phenomenon  differs from the previously reported beam self-cleaning effect associated with a single filament, in the sense that a fundamental mode does not necessarily emerge after self-induced symmetrization. 
We used two indicators to distinguish between beam self-cleaning and self-induced symmetrization: the global beam symmetry degree $\Phi$ and the beam quality factor $M^2$. 
For the single filamentation regime, beam self-cleaning reported previously is essentially associated with an improvement of the beam quality factor, even if the  beam symmetry degree may improve as well.
In contrast, in the case of high power beams considered here, only global symmetry improves substantially, whereas the beam quality factor always remains far from that of a Gaussian beam or the Townes mode ($M^2\sim1$). 
To illustrate this effect, we performed experiments and numerical simulations on the nonlinear propagation of heavily distorted input beams characterized by a dramatic decrease of their symmetry degree after passing through intensity masks. Results clearly demonstrate that the combination of the Kerr effect and focusing is crucial to observe self-symmetrization of highly asymmetric beams.

Self-symmetrization of powerful (multiterawatt) laser beams may prove useful when it is important to homogeneously illuminate a target placed in the focal region. This effect also allows us to obtain a relatively uniform cylindrical plasma channel around focus which shape is independent from the quality of the initial laser beam profile. Such property is particularly interesting for applications based on laser filamentation at high power such as guiding of electric discharges \cite{LaFontaineI3ETPS99,LaFontaineJAP2000,BreletAPL12} and contact-less capture of currents \cite{HouardAPL07}, control of aerodynamic flows \cite{LeonovPP2012}, or lasing effect in air \cite{LuoAPB03,DogariuSCI11,SprangleAPL11,LiuOE13,KartashovPRA13}.

\section{\textbf{Experimental and numerical procedures}}

\subsection{Experiments} 
The collimated output $50$ fs pulse from a linearly polarized multi-Terawatt Ti:Sa CPA laser (Enstamobile) at $\lambda_0=800$ nm is weakly focused with an $f = 5$ m lens in air at atmospheric pressure. Burn patterns from the laser beam are recorded on calibrated presolarized Kodak photographic plates at different distances from the focusing lens. This simple technique reliably detects in a single shot the stage of evolution of multiple filaments formed out of a single laser beam \cite{MechainOC05}. Plasma channels give rise to characteristic $\sim 50-100\ \mu$m wide circular burns on the photographic plate, which are easily distinguished from the $\sim1$ mm circular dark spots from bright channels and the less intense energy reservoir.
Similar measurements were performed by intentionally adding chirp to the pulse to adjust the peak intensity. The same self-symmetrization effect of multiple filaments was observed for the transform limited $50$ fs pulses and for pre-chirped pulses (see figure captions). This allows us to dismiss temporal effects as the origin of symmetrization.

\subsection{Beam characterization}
For each recorded beam intensity profile, we computed the \textit{center of mass} (CM) on the $\textsc{xy}$- (transverse) plane and then considered the radial intensity traces (with fixed polar angle $\phi$), $I_\phi(r)$ ($r=0$ at the CM). We define the degree of symmetry, able to resolve the internal structure (intensity fluctuations) of the beam profiles, as the weighted average of the intensity fluctuations: $0\leq\Phi\leq1$ (see appendix)


\begin{equation}
\Phi\equiv1-\frac{1}{2\pi^2}\int_{0}^{2\pi}d\phi_1\int_{0}^{\phi_1}d\phi_2\left\vert\frac{\int_r\{I_{\phi_1}(r)-I_{\phi_2}(r)\}}{\int_r\{I_{\phi_1}(r)+I_{\phi_2}(r)\}}\right\vert\label{eq2},
\end{equation}
which measures the similarity in between all pairs of the radial intensity traces, $\{I_{\phi}(r),I_{\phi'}(r)\}$.  Perfect circularly symmetric profiles are characterized by $\Phi=1$, whereas beams with a large asymmetry are characterized by a relatively low symmetry degree.

Regarding the beam quality factor, $M^2$, 
we use the widely used definition by Potemkin et al. \cite{PotemkinQE05} for a beam with complex field ${\cal E}(x,y)\equiv A(x,y) \exp(i\phi)$ and intensity $I(x,y)\equiv |{\cal E}(x,y)|^2$:
\begin{equation}
M^2= \left (\langle r^2 \rangle \langle k^2 \rangle - \langle \mathbf{r} \cdot \mathbf{k} \rangle^2\right )^{1/2}
\end{equation}
where
\begin{eqnarray}
P&=&\int dx dy \, I(x,y) \\
\langle r^2 \rangle&=& P^{-1}\int dx dy \,(x^2+y^2)I(x,y) \\
\langle k^2 \rangle&=& P^{-1} \int dx dy  \,[(\nabla A)^2+ I ( \nabla \phi)^2]\\
\langle \mathbf{r} \cdot \mathbf{k} \rangle&=& P^{-1}\int dx dy \, [ {\mathbf r} \cdot \nabla \phi(x,y) ]I(x,y) 
\end{eqnarray}
and $\mathbf{r}\equiv (x,y)$. Since the calculation of the quality factor requires knowledge not only of the beam intensity but also of the spatial phase, we characterized beams in terms of $M^2$ only for the results of our numerical simulations.

\subsection{Simulations}
Propagation of the monochromatic field envelope ${\cal E}(x,y,z)$ in air ($n_0\approx1$), with frequency $\omega_0=2\pi c/\lambda_0$, is modeled by means of a unidirectional beam propagation equation accounting for diffraction, optical Kerr effect, multiphoton absorption, plasma absorption, and plasma defocusing, respectively \cite{CouaironEPJST}:
\begin{multline}
\frac{\partial {\cal E}}{\partial z}=\\
\frac{i}{2 k_0}\left (\frac{\partial^2}{x^2}+\frac{\partial^2}{y^2}\right ) {\cal E} + i\frac{\omega_0}{c}n_2 \vert {\cal E}\vert^2{\cal E}-\frac{1}{2}\left (
\beta_8 \vert {\cal E}\vert^{14}+\sigma [1+i\omega_0\tau_c]\rho 
\right ){\cal E}\label{eq1}.
\end{multline}
Here $n_2=2\times10^{-19}$ cm$^2$/W, $\beta_8=8\times10^{-98}$ cm$^{13}$W$^{-7}$, $\sigma=5.6\times10^{-20}$ cm$^2$, and $\tau_c\approx350$ fs. Our $(2+1)D$ modeling does not contain temporal dynamics. Hence, plasma effects are accounted for in the so-called {\it frozen time} approximation where the explicit standard rate equations describing electron generation by optical field ionization and avalanche are used to generate a mapping of the electric field peak intensity ${\cal{I}}\equiv\vert {\cal E}\vert^2$ to the electron-plasma density, $\rho({\cal{I}})$. This calculation is performed for \textit{reference} pre-chirped Gaussian pulses with peak intensity ${\cal{I}}$ with duration given by experimental conditions. The electron density, $\rho({\cal{I}})$, used in the propagation model Eq. (\ref{eq1}) is determined from this mapping, at the temporal center of the reference pulse.  This procedure provides, as shown previously and below, numerical results in good 
agreement with experiments \cite{HouardPRA08,KiranPRA10,PointPRL14}. Simulations are initialized with $10\%$ intensity and $0.2\%$  phase noise in order to mimic experimental irregularities in the input beam.
\section{\textbf{Results}}
\begin{figure}
\includegraphics[scale=.23]{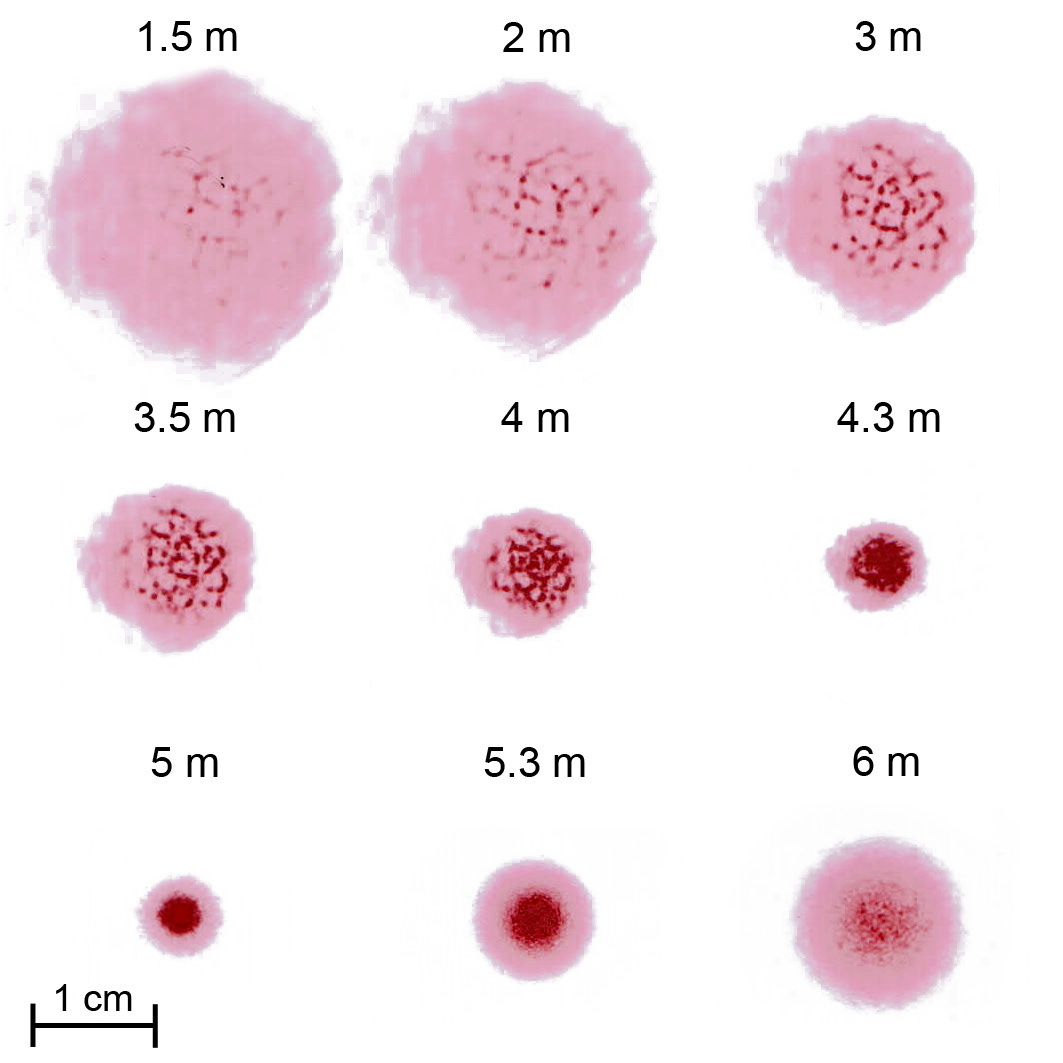}
\caption{Experimental transverse profiles of an undistorted focused beam. Images recorded on the photographic papers at various distances for input pulses of $50$ fs (not pre-chirped), carrying $130$ mJ each and having a beam width of $\sim3.5$ cm. The lens has its geometrical focus at $5$ m. Beam and pulse widths are given at the intensity FWHM and propagation distances are marked on top of each image. The $1$ cm length bar is included for reference. \label{f1}}
\end{figure}
\subsection{Undistorted beams} 
In the absence of any mask, a beam with power $P\sim550\ P_{cr}$ undergoes typical multi-filamentation dynamics. Figure \ref{f1} shows the darkening pattern impressed by such a beam at several distances from the focusing lens on photographic plates. Already after $2$ m of propagation several ionizing filaments are present. These converging filaments grow in number and connect into networks upon further propagation ($\gtrsim4$ m). Before the geometric focus, the ionized zone fills a central circular spot of a few mm in diameter and $\sim1$ m long, where filaments are in close contact \cite{PointPRL14}. The number of ionized spots is significantly reduced to a few which are located near the center of the beam. Losses incurred by going through the focus have been measured with a calorimeter placed before and after the focus.  $82\%$ of the initial beam energy is found in the beam at $3.5$ m after the focus. This rather high value could be connected to the idea that in certain situations the filaments are connected stronger to plasma defocusing than to plasma absorption \cite{MlejnekPRL99}. Below we show that independently on the intensity mask used to distort the input beam profile, the beam recovers after focus the same symmetry degree, hence similar roundness, as the one shown in the final stage of Fig. 1. In contrast, low energy beams ($P\ll P_{cr}$) do not exhibit any improvement of their symmetry degree along propagation since the final beam profile corresponds to the diffraction pattern induced by the mask, with identical symmetry degree as the initial distorted beam. 
\begin{figure}
\includegraphics[scale=.23]{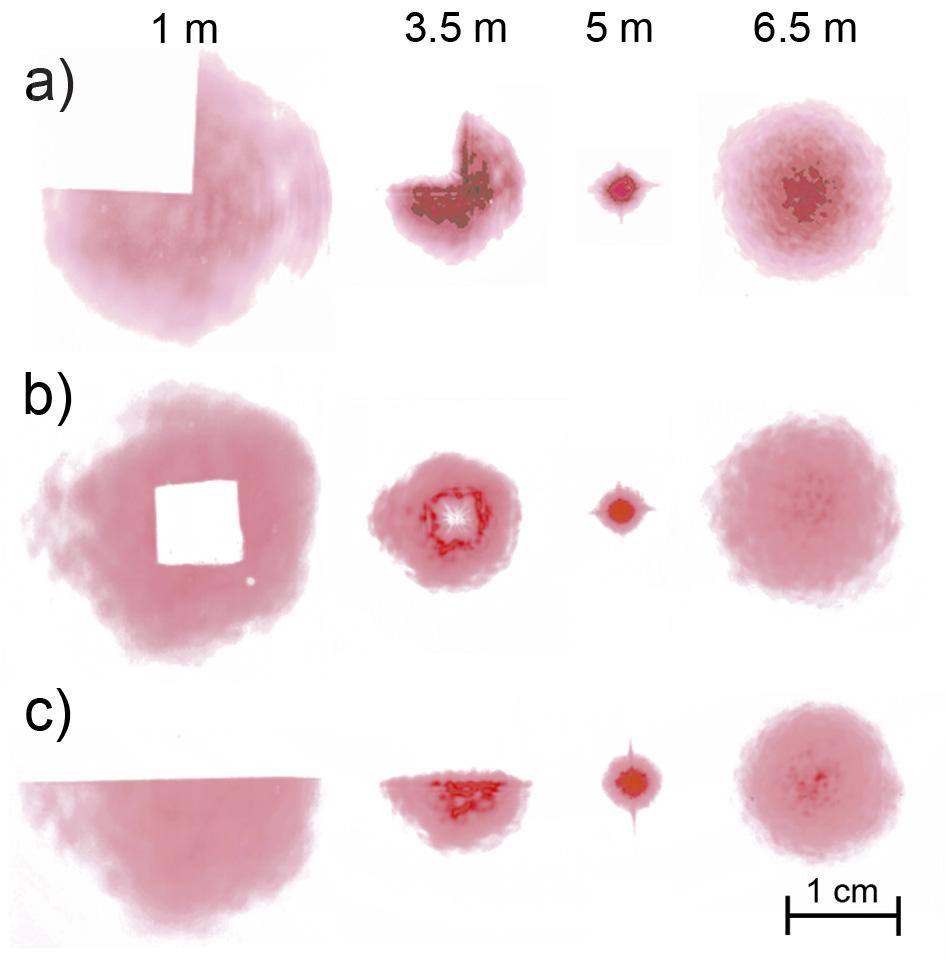}
\caption{Experimental beam profile evolution under spatial distortion of the pre-chirped $700$ fs pulses carrying $295$ mJ before the masks ($P/P_{cr}\approx10^2$). The different masks screen (a) 25 $\%$, (b) 10 $\%$, and (c) 50 $\%$ of the input energy. Input beam widths and focusing conditions are those of Fig. \ref{f1}. Propagation distances are indicated on top of each column of images. \label{f2}}
\end{figure}

\subsection{Self induced symmetrization of highly distorted beams} 
\begin{figure}
\includegraphics[scale=.11]{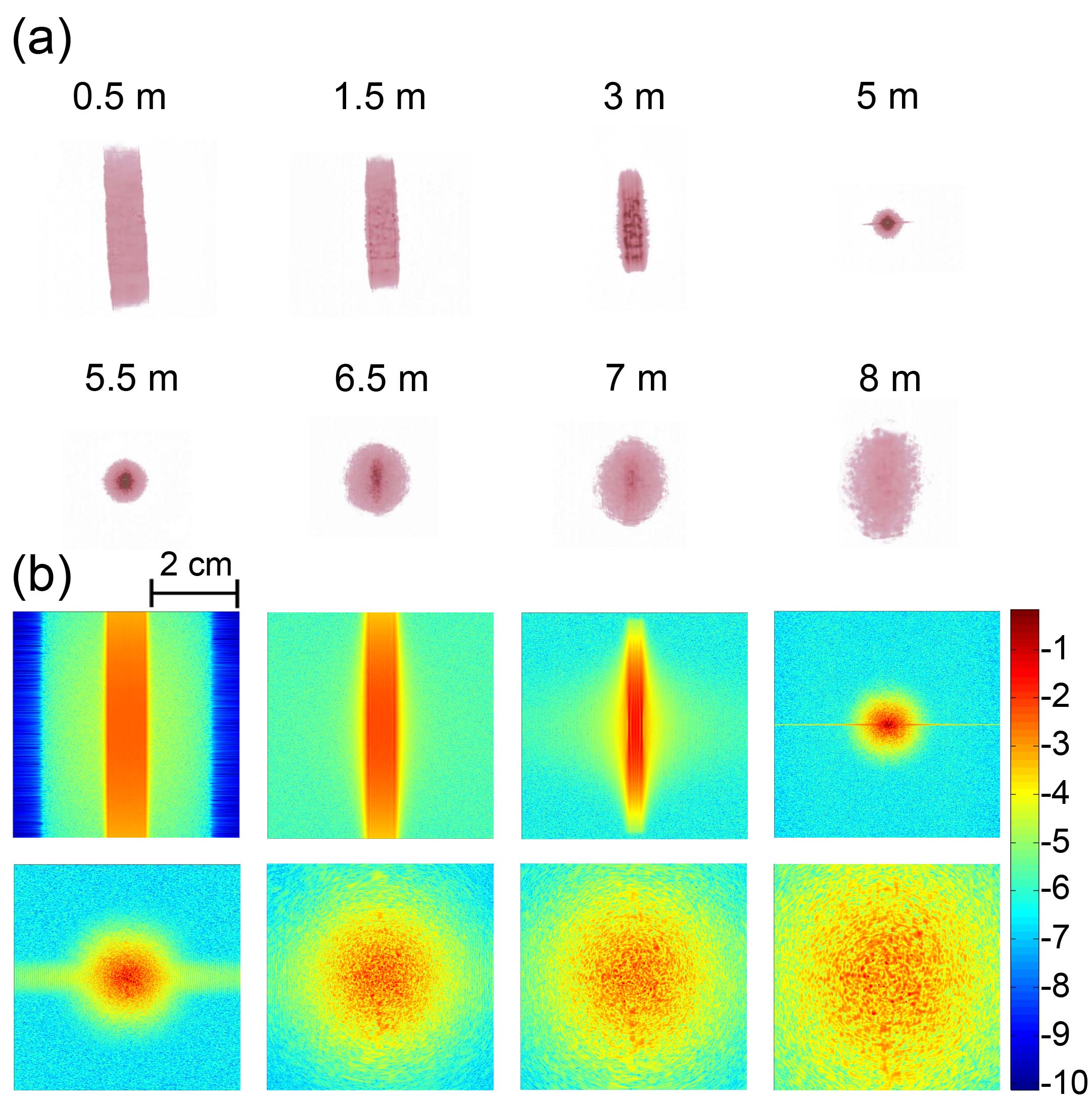}
\caption{Beam evolution after distorting the input beam with a 1 cm wide slit mask, screening $\sim64\ \%$ of the input energy $\sim300$ mJ. (a) Experimental and (b) simulated beam intensities along propagation at selected distances, marked in (a). Pulse width is $100$ fs (pre-chirped) and energy $\sim100$ mJ: $P/P_{cr}\approx210$ (after mask). Simulations are initialized with a peak intensity of $\approx250$ GW/cm$^2$. Position of geometrical focus is at $5$ m and intensities given by the color bar are in Bels.\label{f3}}
\end{figure}

In order to visualize the self symmetrization effect we have put different masks on the path of the input beam, just before the focusing lens (at $0$ m). Examples are shown in Figs. \ref{f2} and \ref{f3} where a quarter circle (or Pac-man), square, half-plane, and slit masks are used.  Similarly to the undistorted case, $89\%$ of the input chirped pulse energy was measured $3$ m beyond the focus. Inspection of the burnt papers in Figs.~\ref{f2} and \ref{f3}(a) reveals that ionizing filaments get first organized along patterns dictated by the profile of the input beam \cite{MechainPRL04}. Indeed, close to the sharp intensity jumps created by the mask, large intensity modulations drive filamentation in a deterministic manner. Upon further beam propagation, the self-symmetrization effect appears around the geometric focus symmetrizing the slowly diffracting output beam, despite its initial strong distortion. Solid experimental evidence of the symmetrization effect is provided by comparing the profiles in Fig. 2 recorded at 1.5 m before and after the focus (i.e., at 3.5 and 6.5 m).

\begin{figure*}
\includegraphics[scale=.33]{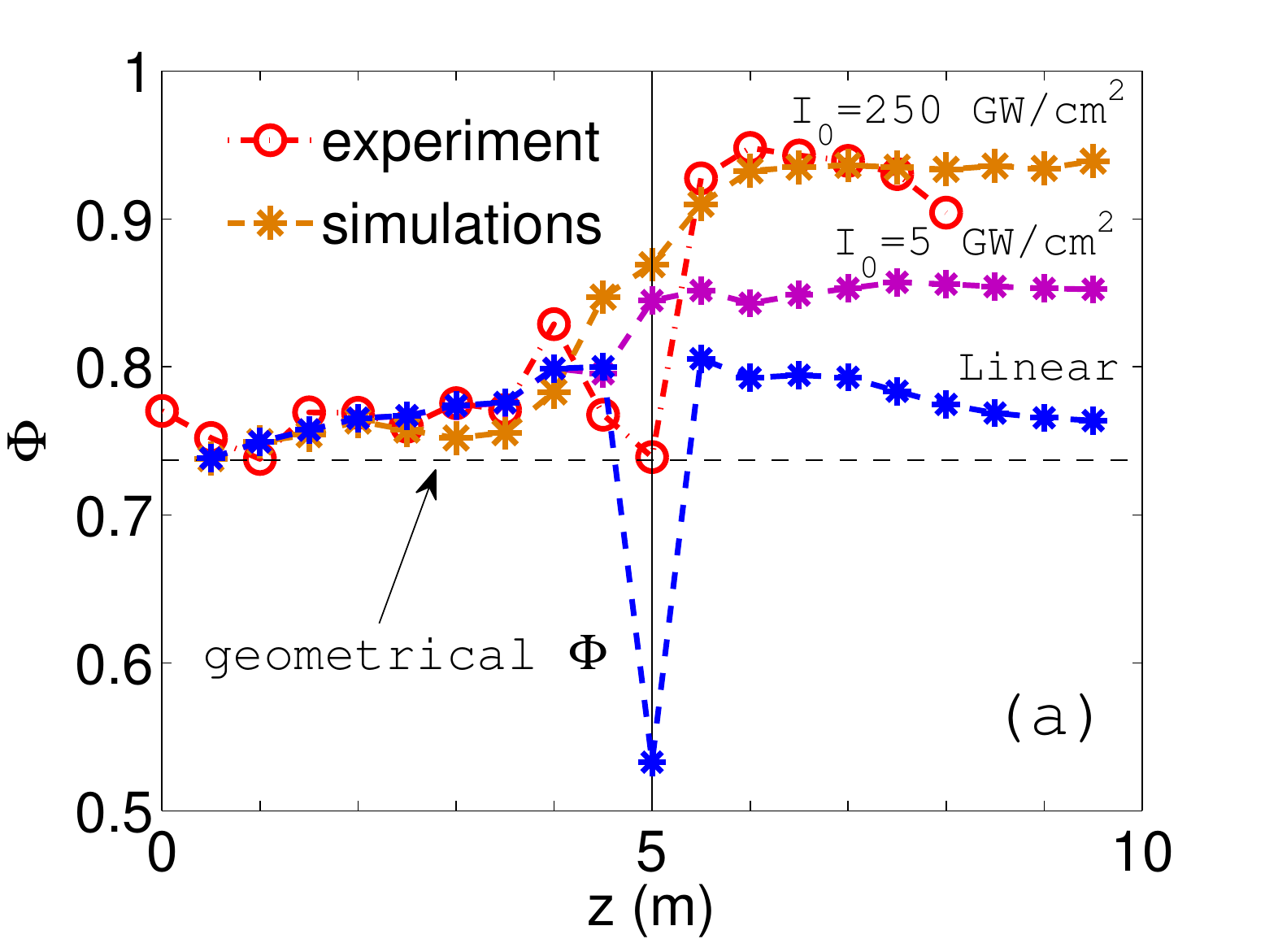}
\includegraphics[scale=.33]{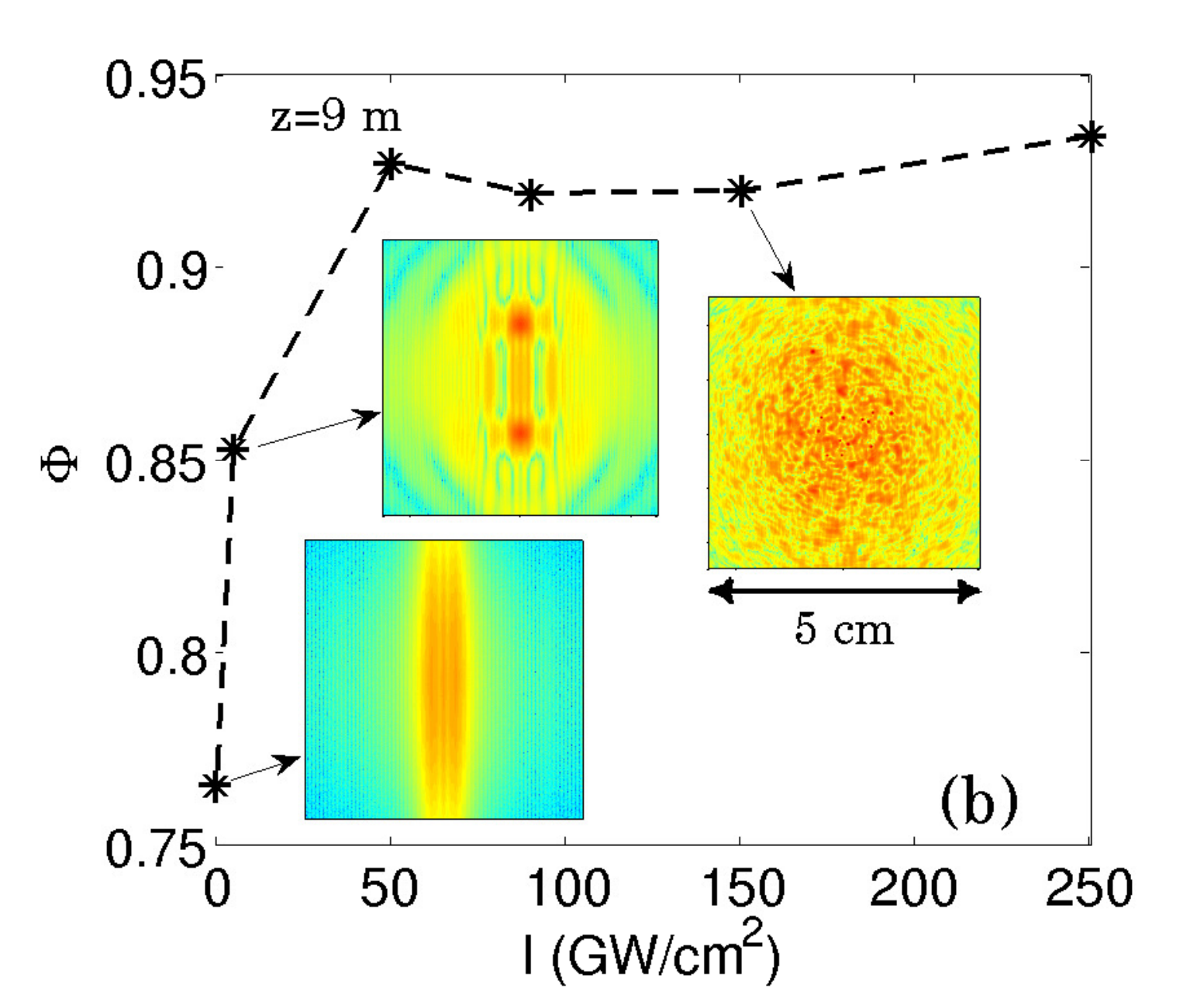}
\includegraphics[scale=.33]{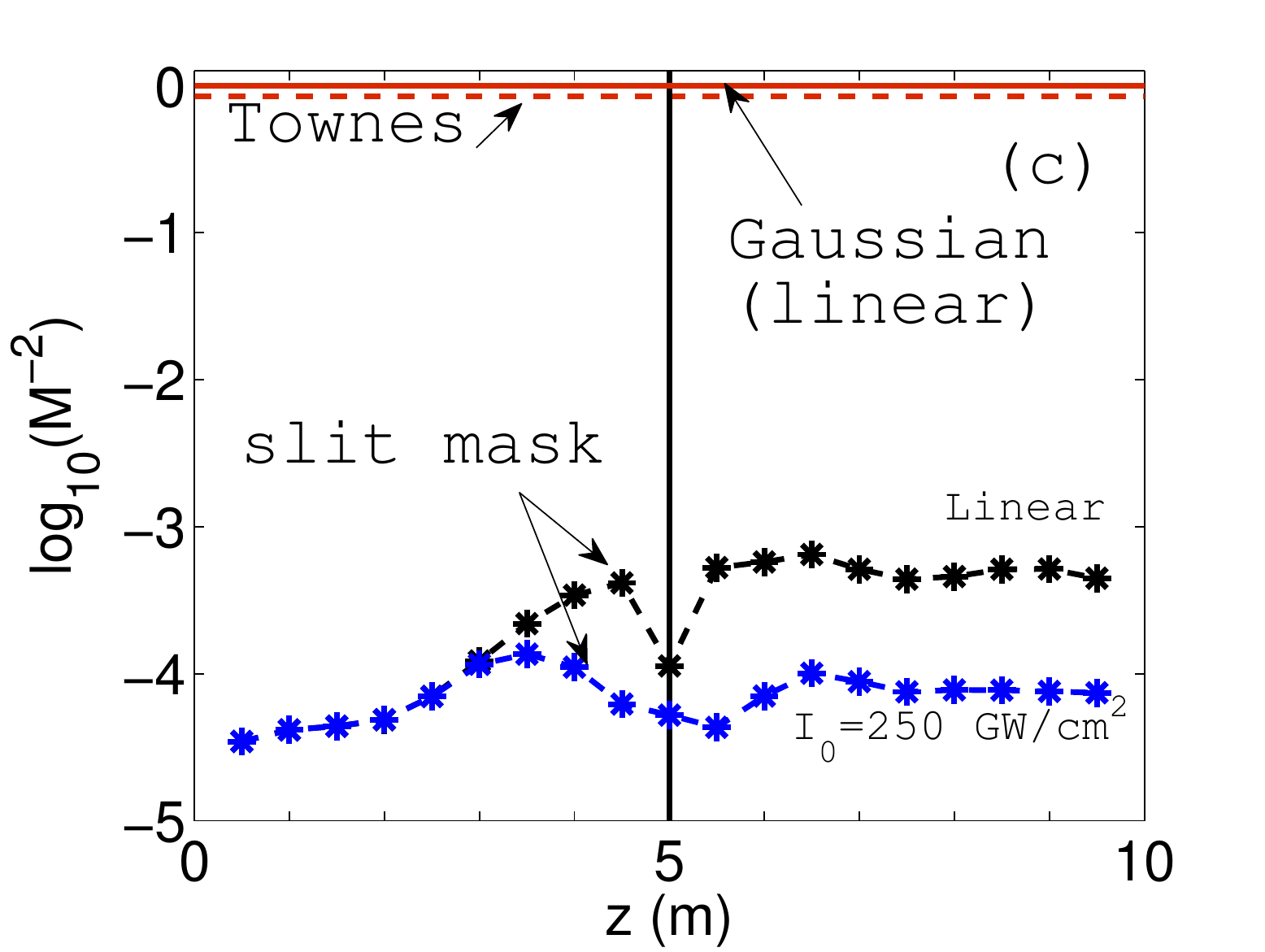}
\caption{(a,b) Symmetry degree of the beam profiles versus propagation and input peak intensity, respectively. (c) Smoothness along propagation for Townes mode (dashed), undistorted focused Gaussian beam propagating in the linear regime (solid-red), and heavily distorted beams, corresponding to the linear and $I_0=250$ GW/cm$^2$ cases shown in (a). Results shown here correspond to the slit mask (see also figures 3 and 5).\label{f4}}
\end{figure*}

A detailed comparison between experimental and numerical results is presented in Fig. \ref{f3} for the case of the highest input distortion. Here, a $10$ mm wide slit  transforms the input circular profile into a rectangle with aspect ratio $\sim1/4$. Upon propagation the symmetrization degree increases as the beam approaches and goes through the focal region at $\sim5$ m. Figure \ref{f4}(a) shows symmetrization of the beam along propagation for the beam profiles presented in Figs. \ref{f3}a-b. For a comparison, two additional numerical results for linear propagation and for an input intensity $I_0=5$ GW/cm$^2$ are also presented. The symmetry degree is seen to be higher by the end of the nonlinear propagations, reaching its maximum after the focus and holding it for a distance of at least $\sim2-3$ m. The minimum of the symmetry is observed for the linear propagation at the focal plane. This is simply due to the sharp horizontal line ($y=0$) apparent in the beam profiles at this position, $z=5$ m (see below, e.g., in Fig. 6), corresponding to the Fourier transform of the slit induced sharp profiles along x (which is a sinc(x) function at focus). One can easily see that the width of the Fourier transforms of a Gaussian, $\Delta\kappa_G$ (at FWHM), and a slit, $\Delta\kappa_s$ (full width of the cetral "sinc" lobe), satisfy $\Delta\kappa_s/\Delta\kappa_G=w/L_s\times\pi/\sqrt{\ln 2}$, where $w\approx3.5$ cm is the FWHM of the input ($z=0$) Gaussian profile along y and $L_s=1$ cm the slit width (along x). It therefore follows that whilst $L_s\lesssim w$, the horizontal line appearing at focus will be much longer than the spot size along y affecting substantially the symmetry degree (see the sharp minimum $\Phi\approx0.5$ in Fig. 4a). This feature disappears gradually as the input peak intensity is increased in simulations up to $I_0\approx5$ GW/cm$^2$, because of the novel spatial frequencies that are generated (see discussion in section V). Also, the photographic paper used in experiments has a different definition in the low intensity parts of the beam and therefore the threshold intensity at which the horizontal line disappears is expected to be different than for simulations. Note in the linear propagations for the masked beams (not shown) the beam profiles at a fixed distance before and after the focus are almost equal to each other after a rotation of $\pi$ in the $\hat{XY}$ plane. The equality does not hold exactly due to the input noise and mask induced fluctuations, however the difference is small at sight. This is the reason why $\Phi(z)$ is (almost) symmetric from the focal plane (see Fig. 4a). In our experiments, the slit mask is the one screening a biggest fraction ($\sim64\%$) of the input 300 mJ carried by laser pulses, which was the maximum available energy. We believe this is why the experimental symmetry degree along propagation (circles in Fig. 4a) starts to drop a bit $2-3$ m after focus. This feature may be appreciated even visually in Fig. 3(a), conversely to what is shown in Fig. 2 for the other masks. Additionally, the symmetry degree presents a local minimum around focus, which is a reminiscence of the linear propagation (indeed we see similar $\Phi(z)$ trends in our monochromatic modeling for $I_0\lesssim10$ GW/cm$^2$). Still, the symmetry degree is remarkably high during 2-3 m and we expect (see below) that this would improve for higher input pulse energies.

The quantitative agreement in between our monochromatic numerical simulations and experimental results in Figs. \ref{f3} and \ref{f4}(a) strongly suggests that the spontaneous symmetrization shown here is an effect essentially dominated by spatial dynamics of the beam. We have computed the symmetry degree at a fixed distance $z=9$ m ($4$ m after focus) and varying input power (Fig. 4(b)). These measurements reveal an increase of $\Phi(P/P_{cr})$ exhibiting saturation: $\Phi(P/P_{cr}\gtrsim40)\approx0.95$. Such behavior vividly manifests the nonlinear (intensity dependent) nature of the spontaneous symmetrization. Indeed, a systematic numerical study reveals that the Kerr effect plays a major role in self-symmetrization (see section \ref{sec-sns} below). We also characterized the beam quality via beam quality factor \cite{SiegmanUS86,PotemkinQE05}. A Gaussian beam at waist is characterized by $M^2=1$ and is considered as a reference for perfect quality. Figure 4(c) shows that the highly distorted beams always present a beam quality $3$ to $5$ orders of magnitude worse than that of a Gaussian beam at focus or the Townes mode. Therefore, even if there is a partial improvement of $M^2$ with propagation distance, the main effect is self-symmetrization. Beam self-cleaning occurs in the sense of an improvement of the beam quality factor but the latter effect is not so efficient as for beam self-cleaning obtained in the context of lower power beams \cite{MollPRL03,PradeOL06,LiuPRA2007}, where $P/P_{cr}\gtrsim1$.

\section{Numerical analysis of key physical effects in beam symmetrization}
\label{sec-sns}
\subsection{Impact of the different nonlinear effects}
We consider below the case of an input beam with a peak intensity of $I_0=50$ GW/cm$^2$ under the distortion of the slit mask. This situation corresponds to the minimum peak intensity needed to observe a high symmetry degree (see Fig. 4(b)) and only a few filaments may be observed on the output beam profile, note there are only four filaments in Fig. 5 (marked by the squares). In order to show that the main responsible for self-symmetrization of powerful focused beams is indeed the optical Kerr effect, we made a set of simulations with identical initial conditions as those in Fig. 5 in which the different nonlinear terms in Eq. (1) are switched \textit{on} or \textit{off} at will (see Fig. 6).  Below, we are only showing beam profiles at and after the focus ($z\geq5$) because those before the focus are indistinguishable by sight from those in Fig. 3 (b) for all cases.

\begin{figure}
\includegraphics[scale=.23]{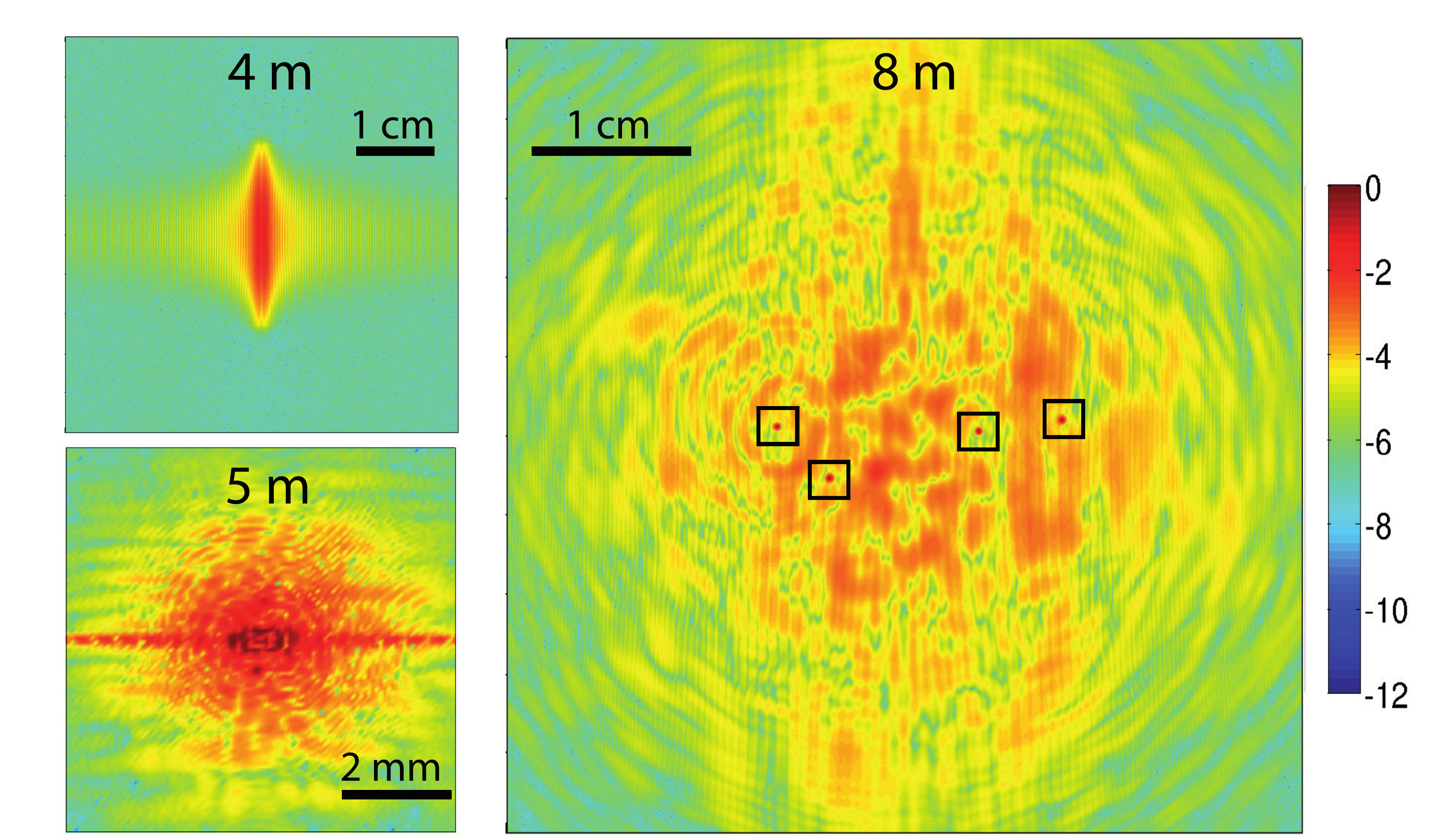}
\caption{Beam symmtrization in the low filament limit. Logarithmic scale plots of beam Intensity profiles at various propagation distances (see labels) for a beam focused with a lens of $5$ m of focal length and perturbed with a $10\times35$ mm$^2$ slit mask (as in Figs. 3 and 4). Input peak intensity $I_0=50$ GW/cm$^2$. All nonlinear effects of Eq. 1 are switched on ($n_2$, $\beta_8$, $\sigma\neq0$). \label{Af1}}
\end{figure}

\begin{figure}
\includegraphics[scale=.3]{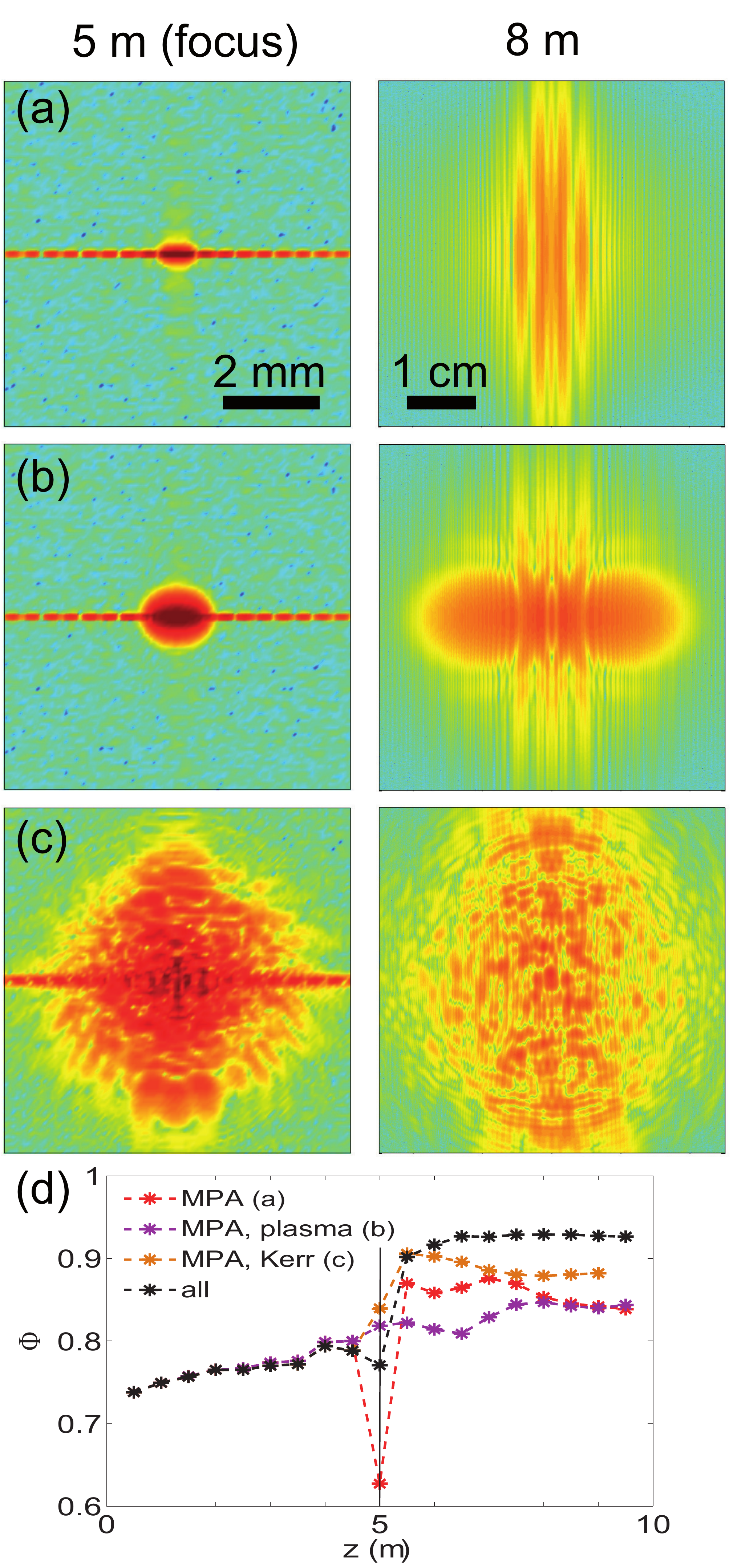}
\caption{Influence of the different nonlinear terms. (a)-(c) Beam profiles at $z=5,8$ m from numerical results corresponding to simulations with identical conditions as in Fig. \ref{Af1}: (a) $n_2,\sigma=0$, (b) $n_2=0$, (c) $\sigma=0$. Color scale is that of Fig. \ref{Af1}. (d) Evolution of symmetry degree for (a)-(c) and Fig. \ref{Af1}. The nonlinear effects that are \textit{on} in each simulation are listed in the legend.  \label{Af2}}
\end{figure}

When the only nonlinear effect is MPA ($n_2$, $\sigma=0$, Fig. \ref{Af2}(a)), propagation does not differ substantially from the one observed in the linear regime, except for the big absorption induced across the profile that lead to the appearance of vertical fringes. Linear propagation would lead to the observation of the diffraction pattern of the slit. Nonlinear absorption plays the role of a distributed stopper, leading to modulations in the diffraction pattern in an effect similar to the Arago spot effect. Nonlinear absorption participates to the fringe formation since it enhances the self-healing process of a beam that is known to reshape Gaussian beams into Bessel-like beams \cite{FaccioOE08b,Porras08OE,Gaizauskas_SF2009}. Addition of plasma absorption and defocusing ($\sigma\neq0$, Fig. \ref{Af2}(b)) has a strong impact in the profile after focus: generation of plasma leads to refractive index jumps along the sharp intensity edges and the defocusing induces strong scattering towards the directions perpendicular to those edges (light propagates towards the higher index regions). 
As a consequence of this, the scattering of light after focus occurs predominantly at an angle $\pi/2$ from the long axis of the slit. It is only when the Kerr effect is switched on that symmetrization degree improves substantially, as shown in Fig. \ref{Af2}(c) ($\beta_8=0$) and Fig. \ref{Af1}. Note from Fig. \ref{Af2}(d) that with Kerr effect symmetry is higher when both MPA and plasma effects are accounted for, presumably due to the fact that MPA and plasma tend to induce scattering along perpendicular directions (at least in the case of the slit mask). Remarkably, the symmetrization effects persists even for beams experiencing a large multifilamentation (see Fig. 3).

\begin{figure}
\includegraphics[scale=.3]{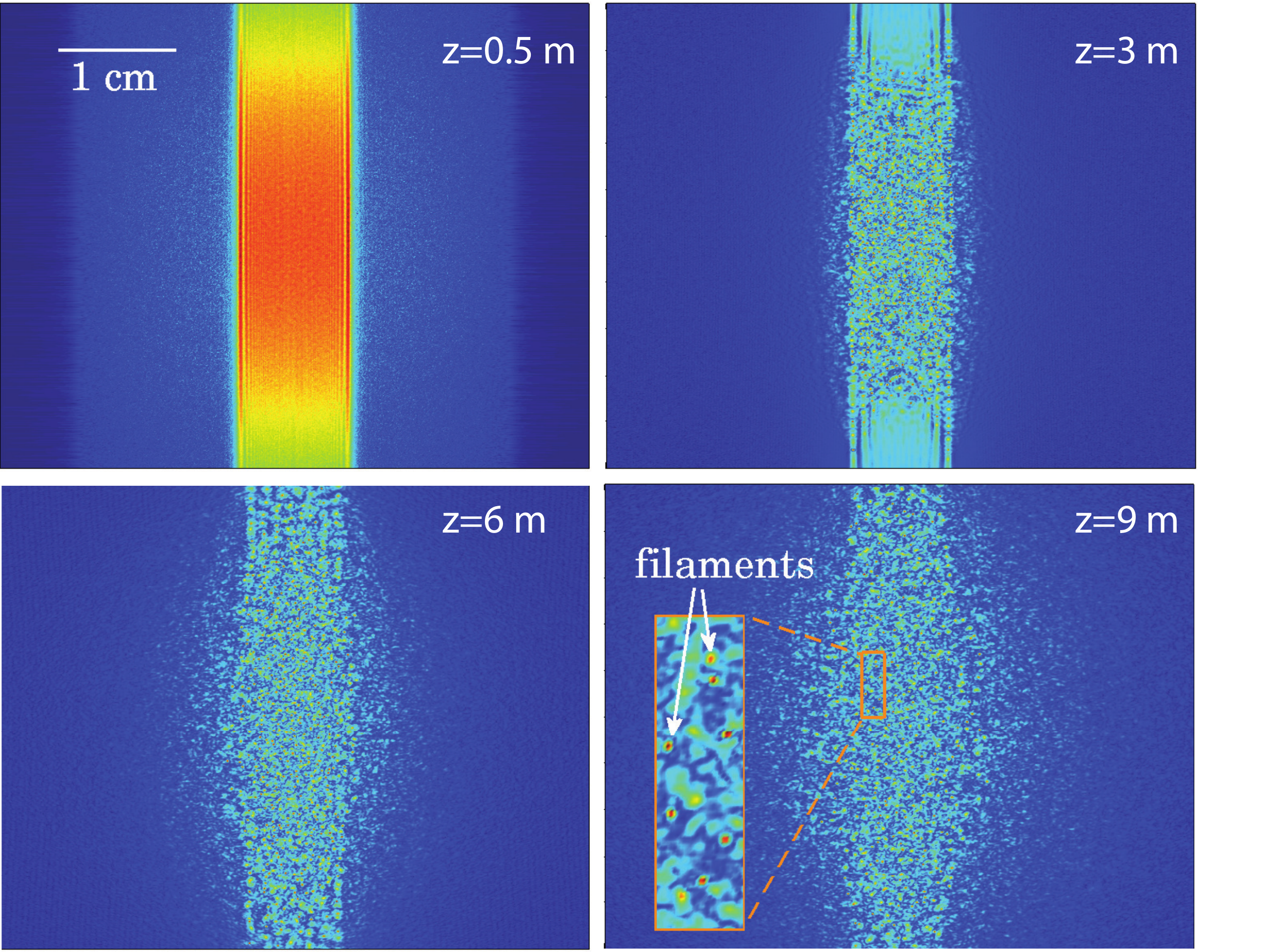}
\caption{Simulated beam intensity cross sections along propagation for a collimated cw beam under distortion of the slit mask. Absence of beam symmetrization and of filament density increase are evident (compare with Fig. \ref{f3}(b)). Numbers in figures mark propagated distances along $z$. The reference Gaussian pulse used in this modeling is of $100$ fs and the input beam peak intensity is $2.5$ TW/cm$^2$. All figures have a cross-section of $4\times4$ cm$^2$.\label{f6}}
\end{figure}

\subsection{Impact of diffraction: non focused beams} 
The second essential ingredient for symmetrization (together with Kerr) is the focusing induced by the lens. We have done extensive numerical modeling with collimated beams and controlled the density of filaments via the input power, $P$. None of these simulations showed traces of spontaneous symmetrization in the distances of $\sim 10$ m. Figure \ref{f6} shows an example with the slit mask. Here, in order to decrease filament separation, the input power was $4$ times larger than for the focused cases shown in Fig. 3(b) or Fig. 4 with the largest $I$. The multiple filaments develop along propagation and the phase gradients impinged by the mask make some of the optical energy migrate into the initially unpopulated area. However, symmetry is seen to remain rather low along the $10$ m, which implies that the self-symmetrizing effect reported here is intimately related to the focusing induced interaction amongst the different parts of the beam.

\section{\textbf{Discussion: a possible explanation for the symmetrization effect}}
In this section we attempt to give a plausible explanation for the symmetrization effect on the basis on the results reported and described above. First of all we notice that results in Sec. IV show that symmetrization occurs mainly due to the Kerr term and that it is not necessarily linked to the presence of many filaments in the beam profile. This strongly suggests that symmetrization occurs as a consequence of the Kerr-induced generation and annihilation of spatial frequencies (this is indeed a necessary condition). Along this line it is well known (see e.g., \cite{Agrawal_book}, for the description in time domain) that the Kerr term induces generation of spatial frequencies more efficiently in the regions where intensity gradients are bigger (e.g., sharp edges impinged by the mask on the beam). The transverse wavenumber of these waves satisfies the proportionality relation $\kappa_\perp\propto\partial_\perp I$ ($\partial_\perp$ denotes spatial derivative in some direction on the transverse plane $\hat{XY}$). Note here two things: (i) the new spectral component is generated at a sharp edge and travels inwards towards the intense part of the beam, and (ii) this wave being generated, implies that other frequencies present at that edge have been annihilated, i.e., the Kerr effect boosts energy from the strongly divergent spatial frequencies at an edge to new frequencies traveling in the opposite direction over the transverse plane (this is of course equivalent to the qualitative argument explaining why the Towns mode \cite{ChiaoPRL64} exists). In the case of the slit mask, the image described above plays a central role along x (horizontal) and not so much along y (vertical) because the Gaussian profile is much smoother than the rectangular shape. As a consequence of this, one could say that Kerr acts as a moderator which kills waves with high $|\kappa_\perp|$ and therefore symmetrizes the spectral distribution. However, this is not enough to explain the effect reported in this paper. We believe that the second necessary condition for symmetrization is the presence of the external focusing. Not only because it enhances intensity levels and the efficiency of the Kerr effect, but also because it tends to localize all the frequency components of the beam into a reduced focal region that acts as a \textit{quasi}-point source with symmetric spectrum in the transverse plane. This unavoidably leads to an after-focus beam that exhibits a substantially high overall roundness and we can therefore say that the beam self-symmetrized. A precise answer to why this effect still happens in the multi filamentation regime ($P\gtrsim150\ P_{cr}$) and what is the role of the turbulent, by nature, multi filament dynamics \cite{MlejnekPRL99} requires further work and understanding.

\section{\textbf{Conclusion}}
We have demonstrated a self-symmetrization effect in air occurring for high power laser beams experiencing multifilamentation and the action of an \textit{external} focusing force. The collective organization process is very different from self-cleaning and is linked to the Kerr induced isotropic wavenumber redistribution leading to beams with improved overall symmetry. Simulations are in excellent qualitative and quantitative agreement with the observed features of this effect induced by focusing powerful beams. The absence of symmetrization for collimated beams highlights the importance of focusing in the global organization in the form of circularly symmetric beam. A possible explanation for this effect is also provided.

\section*{\textbf{Acknowledgments}}
Authors acknowledge financial support from the French Direction G\'{e}n\'{e}rale de l'Armement (DGA).

\section*{\textbf{Appendix: symmetry degree}}

First of all, for a given 2D beam profile, we locate the center of intensity (center of mass)
\begin{equation}
(x,y)_{CM}\equiv\frac{\int_{xy}(x\vec{u}_x+y\vec{u}_y)I(x,y)}{\int_{xy}I(x,y)}
\end{equation}
at which we locate the origin of the polar coordinates, (r,$\phi$): $r=0$ at $(x,y)_{CM}$. Then we consider the radial intensity traces at a fixed polar angle, $I_\phi(r)$, and then we compare them by pairs. Given a pair of these traces we define the \textit{local} relative difference

\begin{equation}
\bar{D}(r,\phi_1,\phi_2)\equiv\frac{I_{\phi_1}(r)-I_{\phi_2}(r)}{I_{\phi_1}(r)+I_{\phi_2}(r)}=\frac{I_{\phi_1}(r)-I_{\phi_2}(r)}{2\langle I_{\phi_1,\phi_2}(r)\rangle}\in[-1,1].
\end{equation}
Since $\bar D$ values are more significant in regions where the average intensity, $\langle I_{\phi_1,\phi_2}(r)\rangle$, is locally high, we define the \textit{weighted} average relative difference in between the pair of traces:

\begin{equation}
w_{\bar{D}}(\phi_1,\phi_2)\equiv\left\vert\frac{\int_r\langle I\rangle\bar{D}}{\int_r\langle I\rangle}\right\vert\in[0,1],
\end{equation}
which nullifies for equal intensity traces and tends to 1 for very dissimilar ones. The overall symmetry degree is obtained by comparing all possible pairs of intensity traces:

\begin{equation}
\Phi\equiv1-\frac{1}{2\pi^2}\int_0^{2\pi}d\phi_1\int_0^{\phi_1}d\phi_2w_{\bar{D}}(\phi_1,\phi_2)\in[0,1],
\end{equation}
where the meaning of $\Phi$ is reversed from that of $w_{\bar{D}}$, i.e., $\Phi$ is good (bad) for values close to 1 (0) and the normalization factor is $\int_0^{2\pi}d\phi_1\int_0^{\phi_1}d\phi_2=2\pi^2$. Eqns. (9-11) trivially combine to give the symmetry degree presented in Eq. (1). In the numerical analysis we have used the discrete version of the above equation taking into account only a finite number of intensity traces ($\phi_{1,2}\rightarrow m,n$), which adopts the form:

\begin{equation}
\Phi\equiv1-\frac{2}{M[M-1]}\sum_{m=1}^{M-1}\sum_{n=m+1}^Mw_{\bar{D}}(m,n)\in[0,1].
\end{equation}

Note that because $\bar D$ can take positive and negative values, $w_{\bar D}$ might be in principle nullified not only by identical intensity traces, but also for those which are different but cross each other one or more times. By doing so we allow a pair of similar traces to be regarded as equal (i.e., tolerance in this case is better than if absolute value bars are introduced in Eq. (9)) but we risk that two very different traces are taken as equal (we therefore assume the probability for this to be small). This could be regarded as a drawback of this similarity measure and its significance depends in reality on the nature of the analyzed data. However, the absolute value bars in Eq. (10) ensure that little non-zero values of $\omega_{\bar D}$ will accumulate when performing the angular integrals (or sums), worsening the overall symmetry. To finalize, we stress here that we tried six different ways (not specified here) of obtaining symmetry indices and the one presented here proofed (unlike all others) to give systematically a good qualitative correspondence for all situations with different masks and input peak intensities. We recall that the purpose of $\Phi$ is the one of obtaining a global feature of a beam profile with complex structure.



\end{document}